\documentclass[useAMS,usenatbib,asm]{mn2e}
 \usepackage{graphicx,fleqn,times}
\title[Star clusters towards the Galactic Center]{UBVI CCD photometry and star counts in nine
inner disk Galactic star clusters\thanks{Based on observations collected at the Cerro Tololo Inter-American Observatory
and Las Campanas Observatory, Chile.}}

\author[Giovanni Carraro  and Anton F. Seleznev]{Giovanni Carraro$^{1}$\thanks{On
leave from Dipartimento di Astronomia,
Universit\'a di Padova, Italy}\thanks{E-mail:
gcarraro@eso.org (GC); anton.seleznev@usu.ru (AFS)}
and Anton F. Seleznev$^{2}$\thanks{ESO Visiting Scientist}\\
$^{1}$European Southern Observatory, Alonso de Cordova 3107,
Casilla 19001, Santiago 19, Chile\\
$^{2}$Astronomical Observatory, Ural State University, Lenin Avenue. 51, Ekaterinburg 620083, Russia}

\begin{document}

\date{Accepted 2011..... Received 2011...; in original form 2011...}

\pagerange{\pageref{firstpage}--\pageref{lastpage}} \pubyear{2011}

\maketitle

\label{firstpage}

\begin{abstract}
We present and discuss new CCD-based photometric material in the UBVI pass-bands for nine Galactic
star clusters located inside the solar ring, for which no CCD data are currently available. 
They are: IC~2714, NGC~4052, ESO131SC09, NGC~5284, NGC~5316, NGC~5715, VdB-Hagen~164,
NGC~6268, and Czernik~38.
The main aim of this study is to establish their nature of real clusters
or random field star enhancements and, when real, estimate their fundamental parameters.
To this aim, we first perform star counts by combining our optical photometry with 2MASS,
and derive cluster sizes and radial density profiles.
The fundamental parameters -
age, reddening and distance-  are then inferred from the analysis
of the star distribution in color-color and color-magnitude diagrams
of only the spatially selected likely members.
Our analysis shows that ESO131SC09, NGC~5284, and VdB-Hagen~164 are most probably not clusters,
but random enhancements of a few bright stars along the line of sight, with properties much similar
to so-called open cluster remnants.
The remaining clusters are physical groups, and are all younger than about 1 Gyr.  We use the newly
derived  set of
parameters, in particular distance and reddening, to investigate their position in the Galaxy in
the context of the spiral structure of the Milky Way.
We find that the youngest clusters (IC~2714, NGC~5316, and NGC~6268)
are located close to or inside the Carina-Sagittarius arm, and are therefore {\it bona fide}
spiral structure tracers.
On the other hand, the oldest (Czernik~38, NGC~4052, and NGC~5715) are floating
in the inter-arm space between the Carina-Sagittarius and the more
distant Scutum-Crux arm.
Interestingly enough, 
the oldest clusters of this sample - Czernik~38 and NGC~5715-
are among the few known open clusters to be older or as old as the Hyades in the inner 
Galactic disk, where star clusters are not expected to survive for a long time,
because of the strong tidal field and the higher probability of close encounters.
\end{abstract}


\section{Introduction}
In the Milky Way, stellar clusters form in dense regions located inside spiral arms.
When clusters survive, they remain connected with the parent arm for about 100 Myr
(Dobbs \& Pringle 2010). Afterwards
they decouple from it and are no longer useful as spiral structure tracers.
Young star clusters have been used for half a century to probe the spiral
structure of the Milky Way (MW).
Historically, the first arms to be detected using young open clusters,
were the Perseus arm in the second
galactic quadrant, the Carina-Sagittarius in the fourth quadrant, and the Orion spur
in which the Sun is embedded (Trumpler 1930).
Nowadays, the picture we have of the MW spiral structure contains many
more details (Russeil 2003,
Efremov 2011, Lepine et al. 2011), although a lively discussion is still ongoing as to how
many major arms are present and if they are long-lived or transient
(Grosbol et al. 2011).

In the last decade, young star clusters played a major role in improving our knowledge
of the MW spiral structure, especially in the third Galactic quadrant.
Moitinho et al. (2005), Vazquez et al. (2008) and Carraro et al. (2010) identified
for the first time in optical observations the outer Norma-Cygnus arm, and clarified the shape
and interaction of the Orion and Perseus arms. These studies made clear
that young star clusters are powerful  spiral tracers when it is possible
to determine their distance and age with high confidence.
In particular, the authors stress how crucial deep $U$-band photometry is to pin down
cluster reddening and hence obtain their distance.\\

\noindent
We also present and discuss  $UBVI$ photometry of nine star clusters all
located in the fourth Galactic quadrant except  for one - Czernik~38-, which
is located in the first quadrant (see Table~1).
Most of them are nothing more than simple star cluster candidates,
according to most public catalogs, and do not have any published data. 
Howvere, we expect that they are mostly
young clusters based both on inspection of DSS maps and on the widely spread idea
that clusters cannot survive for a long time in the inner Galactic disk.

Apart from the obvious goal to establish their nature and derive their fundamental
parameters, we aim to use them - when they are sufficiently young-
as spiral arm tracers in the longitude range $290^{o} \leq l \leq 360^{o}$.
In this MW sector, inside the solar ring, we expect the lines
of sight of the clusters to first encounter  the Carina-Sagittarius arm, and hopefully,
also the Scutum-Crux arm, which is much closer to the Galactic Center, as well as the Perseus arm,
which is much further away (V\'azquez et al. 2005; Carraro \& Costa 2009; Baume et al. 2009).\\

The technique we used is based on the analysis of the color-color and color-magnitude diagrams
of groups of stars properly selected after having performed star counts and defined
clusters' reality and size (see Seleznev et al. 2010 for details).\\

\noindent
As a consequence, the paper is organized as follows:\\
In Sect.~2 we present information culled from  literature on the objects under investigation,
which demonstrate they have been almost completely overlooked in the past.
Sect.~3 presents the observations and basic data reduction, together with 
photometric calibration, completeness analysis and cross-correlation
with 2MASS. The comparison with previous photometry - when available- are discussed
in Sections~4. Section~5 is devoted to star counts and the derivation of clusters' size.
We then estimate the method to infer clusters' fundamental parameters in Sect.~6. 
Sect.~7, finally, discusses the outcome of this analysis. 
The conclusions of this investigation, together with suggestions/recommendations
for further studies, are provided in Sect.~8.

\section{Previous investigations}
Most clusters in the sample we investigated for this paper have not been studied so far, and only for
one -Czernik~38- some prelimiary CCD data have been provided.\\
We present here a compilation of any previous study we could find in the literature to date,
on a cluster-by-cluster basis.\\

\noindent
{\bf IC~2714}\\
This is the star cluster of the whole sample which received more attention in the past. 
We selected this cluster also as a control
object for our photometry, which we are going to compare in Sect.~4. \\

\noindent
UBV photo-electric photometry of about 200 stars have been presented by Claria
et al. (1994), together with spectroscopy of 14 probable giant star members of the cluster. These
authors derived an
age of 300 million years and place the cluster at a distance of 1300 pc, in agreement
with previous findings by Becker (1960). They found that the cluster suffers from variable extinction,
and obtain E(B-V)=0.36 from the turnoff, and 0.37 from giant stars.
As for the metallicity, they found a large range of value, from +0.05 using DDO to -0.18 using
Whasington photometric system. A revision of the DDO calibration led Twarog et al (1997)
to provide  [Fe/H] ranging from -0.01 to 0.02, and E(B-V)=0.35.
More recently, spectroscopic analysis by Santos et al. (2009) confirmed the range of metallicity
found by Twarog et al. (1997), ruling out the results of the Washington photometry.
Finally, Smiljanic et al (2009) derived [Fe/H]=0.12 from one giant, and reddening
E(B-V)=0.33. By adopting Shaller et al (1992) isochrones, they fit Claria et al (1994)
photometry, getting an apparent distance modulus (m-M) =11.5, and an age of 400 million
years.\\

\noindent
{\bf NGC~4052}\\
van den Berg \& Hagen (1975) describe NGC~4052 as a a medium richness,
real cluster, mostly composed of  blue stars, and possibly
embedded in some nebulosity. The shallow study of Kharchenko et al. (2005) provides 
preliminary estimates of the cluster parameters
based on photometry and astrometry of less than 20 stars brighter than V$\sim$12. They locate this
250 million years old cluster at 1200 pc from the Sun. No CCD studies of NGC~4052 have been conducted
so far.\\

\noindent
{\bf ESO-131SC09}\\
No information is available for this cluster apart from its size, which is around 5 arcmin
according to Dias et al. (2002).\\

\noindent
{\bf NGC~5284}\\
No information is available for this cluster beyond the mere classification.\\

\noindent
{\bf NGC~5316}\\
The only available study for this cluster is the photo-electric investigation by Ramin (1966)
down to V $\sim$16, whicj
places the cluster at an heliocentric distance of 1200 pc. The age of 150 million is derived by
Kharchenko et al. (2005) from a subsample of stars brighter than V$\sim$12. Its young age is also
recognized by the visual inspection of  van den Berg \& Hagen (1975).\\

\noindent
{\bf NGC~5715}\\
van den Berg \& Hagen (1975) describe NGC~5715 as a medium richness, real cluster, mostly composed of
blue stars. No optical study is available for NGC~5715, whose properties have been however investigated
using 2MASS by Bonatto \& Bica (2007). They report an age of 800 million years and a distance of 1.5
kpc from the Sun.\\

\noindent
{\bf Vdb-Hagen~164}\\
van den Berg \& Hagen (1975) describe this cluster as a very poor ensemble of blue stars. No data
is available for it to date.\\

\noindent
{\bf NGC~6268}\\
Seggewiss (1968) provided some photographic photometry of NGC~6268 down to V $\sim$ 14,
and its analysis led to an helio-centric
distance of 1200 pc.
Later, van den Berg \& Hagen (1975) described it  as a very poor ensemble of blue stars.
The shallower study by Kharchenko et al. (2005) found a somewhat smaller distance
of 1000 pc, and suggests an age around 15 million years, lending support to  
van den Berg \& Hagen's (1975) visual impressions.\\

\noindent
{\bf Czernik~38}\\
Discovered by Czernik (1966), this clusters is described as a relatively rich cluster with a diameter
of 14 arcmin. Schmidt camera CCD photometry has been recently obtained by
Maciejewski  (2008). This author suggests that the cluster 
is at least 1 Gyr old, and places it at an helio-centric distance of 1200 pc.\\

\begin{table*}
\caption{Basic parameters of the clusters under investigation.
Coordinates are for J2000.0. In the last two columns we report the extinction
at infinity and the Observatory where data were taken.}
\begin{tabular}{cccccccc}
\hline
\hline
\multicolumn{1}{c}{Number} &
\multicolumn{1}{c}{Name} &
\multicolumn{1}{c}{$RA$}  &
\multicolumn{1}{c}{$DEC$}  &
\multicolumn{1}{c}{$l$} &
\multicolumn{1}{c}{$b$} &
\multicolumn{1}{c}{E(B-V)}&
\multicolumn{1}{c}{Observatory}\\
\hline
& & {\rm $hh:mm:ss$} & {\rm $^{o}$~:~$^{\prime}$~:~$^{\prime\prime}$} & [deg] & [deg] & mag & \\
\hline
1  & IC 2714       & 11:17:27.0 & -62:44:00.0 & 292.40 & $-$1.799 &  2.77 &  LCO\\
2  & NGC 4052      & 12:01:12.0 & -63:13:12.0 & 297.30 & $-$0.900 &  4.15 &  LCO\\
3  & ESO131SC09    & 12:29:37.4 & -57:52:31.0 & 300.02 & 4.873    &  0.87 & CTIO\\
4  & NGC 5284      & 13:46:29.9 & -59:08:39.0 & 309.95 & 2.975    &  1.25 & CTIO\\
5  & NGC 5316      & 13:53:57.0 & -61:52:06.0 & 310.23 & 0.115    & 12.02 &  LCO\\
6  & NGC 5715      & 14:43:29.0 & -57:34:36.0 & 317.53 & 2.085    &  1.78 &  LCO\\
7  & VdB-Hagen 164 & 14:48:13.7 & -66:20:12.0 & 314.28 & $-$6.070 &  0.32 & CTIO\\
8  & NGC 6268      & 17:02:00.8 & -39:44:18.0 & 346.10 & 1.216    &  5.81 & CTIO\\
9  & Czernik 38    & 18:49:42.0 & +05:52:00.0 &  37.13 & 2.630    &  2.08 & CTIO\\
\hline\hline
\end{tabular}
\end{table*}

\section{Observations}
In this section we describe in details how we conducted observations and reduced
the collected data. Five of the program clusters (ESO131SC09, NGC~5284, VdB-Hagen~164,
NGC~6268, and Czernik~38) were surveyed at CTIO (Cerro Tololo Inter-American
Observatory\footnote{http://www.ctio.noao.edu}), 
while the remaining four (IC~2714, NGC~4052, NGC~5316 and NGC~5715)
were imaged at LCO (Las Campanas Observatory\footnote {http://www.lco.cl}). 
Their Equatorial and Galactic coordinates
for the 2000.0 equinox are reported
in Table~1, together with an estimate of the reddening from FIRB (Schlegel et al. 1998).
The last column indicates
where each cluster was observed. Fig~1, finally, shows DSS images of the nine fields where
the clusters are located.\\
Because of the differences in both telescopes and detectors, we will describe
separately the observations and data reduction of the two clusters' groups.

\begin{figure*}
   \centering
   \includegraphics[width=15truecm]{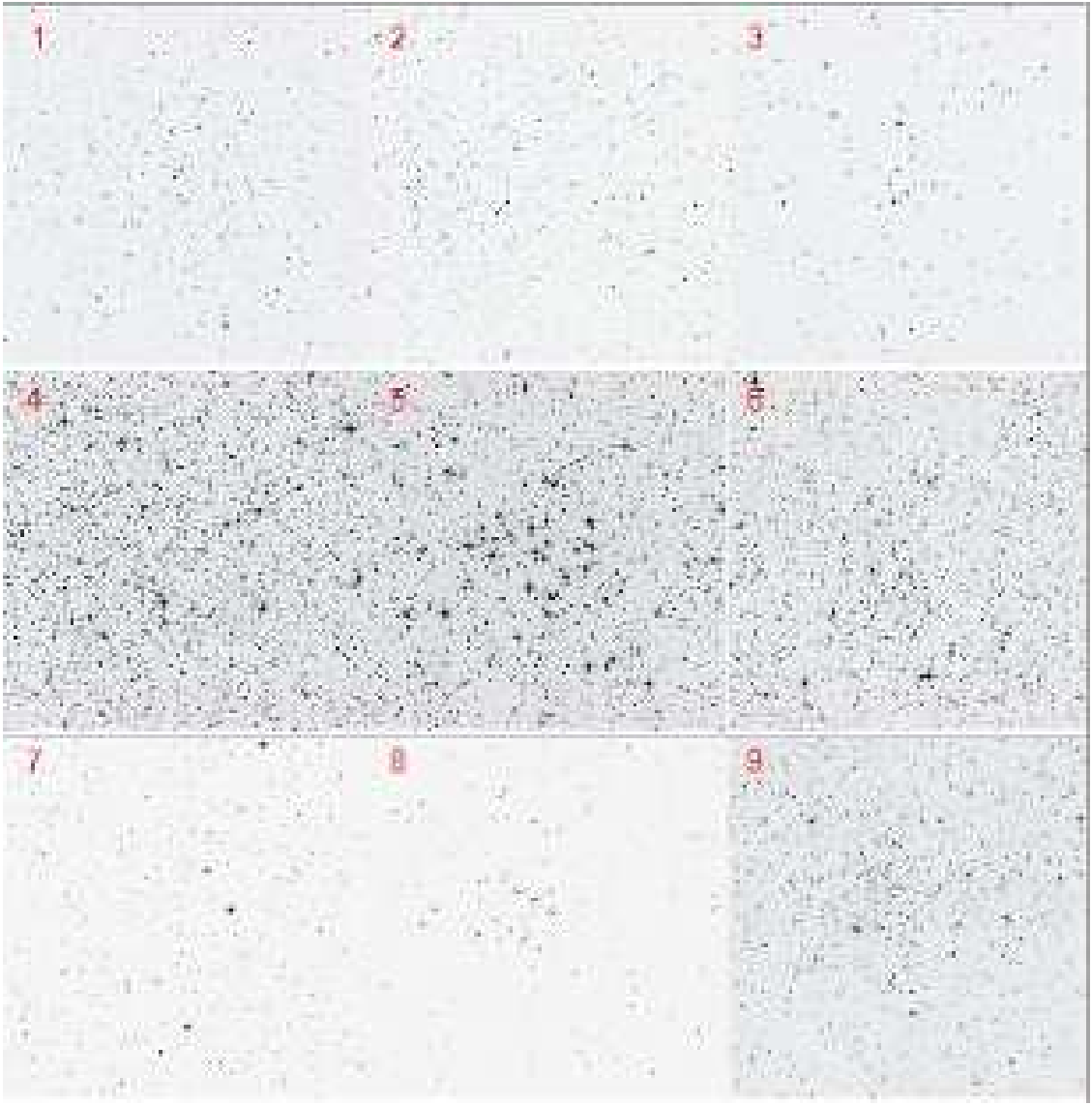}
   \caption{DSS images centered on the 9 program clusters. North is up, East to the left,
   and the field of view is 20 arcmin on a side. Images follow the numbering in Table~1
   from top left to bottom right.}
   \end{figure*}

\subsection{CTIO Observations}
These regions were observed with the Y4KCAM camera attached to the Cerro
Tololo Inter-American Observatory (CTIO) 1-m telescope, operated by the SMARTS
consortium.\footnote{\tt http://http://www.astro.yale.edu/smarts} This camera is equipped with an STA
4064$\times$4064 CCD\footnote{\texttt{http://www.astronomy.ohio-state.edu/Y4KCam/ detector.html}}
with 15-$\mu$m pixels, yielding a scale of 0.289$^{\prime\prime}$/pixel
and a field-of-view (FOV) of $20^{\prime} \times 20^{\prime}$ at the
Cassegrain focus of the CTIO 1-m telescope. The CCD was operated without binning, at a nominal
gain of 1.44 e$^-$/ADU, implying a readout noise of 7~e$^-$ per quadrant (this detector is read
by means of four different amplifiers).
\\
\noindent

In Table~2 we present the log of our \emph{UBVI} observations. All observations were carried out in
photometric, good-seeing (always less than 1.2 arcsec), conditions. Our \emph{UBVI} instrumental photometric system was defined
by the use of a standard broad-band Kitt Peak \emph{UBVI$_{kc}$} set of filters.\footnote{\texttt{http://www.astronomy.ohio-state.edu/Y4KCam/ filters.html}}
To determine the transformation from our instrumental system to the standard Johnson-Kron-Cousins
system, and to correct for extinction, we observed stars in Landolt's areas
PG~1047, PG~1323, G~26 and MarkA (Landolt 1992)
multiple times and with different
air-masses ranging from $\sim1.03$ to $\sim2.0$, and covering quite a large color range
-0.3 $\leq (B-V) \leq$ 2.0 mag.

\begin{table}
%%\centering
\tabcolsep 0.1truecm
\caption{$UBVI$ photometric observations of star clusters and standard star fields for
the CTIO run.}
\begin{tabular}{lcccc}
\hline
\noalign{\smallskip}
Target& Date & Filter & Exposure (sec) & airmass\\
\noalign{\smallskip}
\hline
\noalign{\smallskip}
PG~1047      & 2006 May 21     & \textit{U} & 90,180               &1.15$-$1.25\\
             &                 & \textit{B} & 2x80                 &1.15$-$1.21\\
             &                 & \textit{V} & 2x50                 &1.14$-$1.16\\
             &                 & \textit{I} & 2x50                 &1.15$-$1.15\\
ESO131SC09   & 2006 May 21     & \textit{U} & 30, 200, 1500        &1.13$-$1.13\\
             &                 & \textit{B} & 30, 100, 1200        &1.14$-$1.16\\
             &                 & \textsl{V} & 30, 100, 900         &1.14$-$1.15\\
             &                 & \textsl{I} & 30, 100, 700         &1.13$-$1.14\\
NGC~5284     & 2006 May 21     & \textit{U} & 30, 200, 1500        &1.15$-$1.15\\
             &                 & \textit{B} & 30, 100, 1200        &1.16$-$1.15\\
             &                 & \textit{V} & 30, 100, 900         &1.15$-$1.15\\
             &                 & \textit{I} & 30, 100, 700         &1.15$-$1.15\\
PG~1323      & 2006 May 21     & \textit{U} & 2x150                &1.15$-$1.14\\
             &                 & \textit{B} & 80                   &1.13$-$1.14\\
             &                 & \textit{V} & 50                   &1.15$-$1.15\\
             &                 & \textit{I} & 50                   &1.14$-$1.14\\
VdB-Hagen154 & 2006 May 21     & \textit{U} & 30, 200, 1500        &1.25$-$1.25\\
             &                 & \textit{B} & 30, 100, 1200        &1.26$-$1.25\\
             &                 & \textit{V} & 30, 100, 900         &1.25$-$1.25\\
             &                 & \textit{I} & 30, 100, 700         &1.25$-$1.25\\
NGC~6268     & 2006 May 21     & \textit{U} & 30, 200, 1500        &1.15$-$1.15\\
             &                 & \textit{B} & 30, 100, 1200        &1.16$-$1.15\\
             &                 & \textit{V} & 30, 100, 900         &1.15$-$1.15\\
             &                 & \textit{I} & 30, 100, 700         &1.15$-$1.15\\
MarkA        & 2006 May 21     & \textit{U} & 100, 150, 180        &1.05$-$1.94\\
             &                 & \textit{B} & 3x80                 &1.03$-$1.84\\
             &                 & \textit{V} & 3x50                 &1.05$-$1.85\\
             &                 & \textit{I} & 3x50                 &1.04$-$1.99\\
Czernik~38   & 2006 May 21     & \textit{B} & 30, 100, 1200        &1.46$-$1.45\\
             &                 & \textit{V} & 10, 30,  100, 900    &1.45$-$1.45\\
             &                 & \textit{I} & 30, 100, 700         &1.45$-$1.45\\
G~26         & 2006 May 21     & \textit{U} & 100, 150             &1.35$-$1.74\\
             &                 & \textit{B} & 2x80                 &1.33$-$1.77\\
             &                 & \textit{V} & 2x50                 &1.35$-$1.75\\
             &                 & \textit{I} & 2x50                 &1.34$-$1.79\\
\noalign{\smallskip}
\hline
\end{tabular}
\end{table}

\subsection{LCO Observations}

Four clusters were observed at LCO (see Table~1) on the nights of May 8 to 10,
2010, as illustrated in Table~3. The clusters were observed using the SITe$\#$3 CCD detector
onboard the Swope 1.0m telescope. With a pixel scale of 0.435 arcsec/pixel, this CCD allows
to cover 14.8 $\times$ 22.8 arcmin on sky.
To determine the transformation from our instrumental system to the standard Johnson-Kron-Cousins
system, and to correct for extinction, we observed stars in Landolt's areas
PG~1047, PG~1323, PG~1633, PG 1657, and MarkA (Landolt 1992)
multiple times and with different
air-masses ranging from $\sim1.05$ to $\sim1.9$, and covering quite a large color range
-0.4 $\leq (B-V) \leq$ 2.1 mag (see Table~3). We secured night-dependent calibrations.

\begin{table}
%%\centering
\tabcolsep 0.1truecm
\caption{$UBVI$ photometric observations of star clusters and standard star fields for
the LCO run}
\begin{tabular}{lcccc}
\hline
\noalign{\smallskip}
Target& Date & Filter & Exposure (sec) & airmass\\
\noalign{\smallskip}
\hline
\noalign{\smallskip}
IC~2714      & 2010 May 08     & \textit{U} & 10, 300, 1500        &1.28$-$1.35\\
             &                 & \textit{B} & 10, 200, 1200        &1.25$-$1.26\\
             &                 & \textsl{V} &  3,  30,  900        &1.23$-$1.23\\
             &                 & \textsl{I} &  3,  10,  900        &1.21$-$1.23\\
MarkA        & 2010 May 08     & \textit{U} & 2x240                &1.07$-$1.38\\
             &                 & \textit{B} & 2x180                &1.07$-$1.41\\
             &                 & \textit{V} & 2x60                 &1.07$-$1.43\\
             &                 & \textit{I} & 40,50                &1.07$-$1.43\\
PG~1047      & 2010 May 08     & \textit{U} & 20,180               &1.14$-$1.91\\
             &                 & \textit{B} & 15,180               &1.14$-$1.88\\
             &                 & \textit{V} & 2x10,20,30           &1.15$-$1.83\\
             &                 & \textit{I} & 10,30                &1.15$-$1.86\\
PG~1323      & 2010 May 08     & \textit{U} & 180                  &1.11$-$1.11\\
             &                 & \textit{B} & 120                  &1.12$-$1.12\\
             &                 & \textit{V} & 30                   &1.15$-$1.15\\
             &                 & \textit{I} & 30                   &1.14$-$1.14\\
NGC~4052     & 2010 May 09     & \textit{U} & 10, 300, 1500        &1.28$-$1.35\\
             &                 & \textit{B} & 10, 200, 1200        &1.21$-$1.22\\
             &                 & \textsl{V} &  3,  30,  900        &1.23$-$1.24\\
             &                 & \textsl{I} &  3,  10,  900        &1.22$-$1.22\\
NGC~5316     & 2010 May 09     & \textit{U} & 10, 300, 1500        &1.17$-$1.20\\
             &                 & \textit{B} & 10, 200, 1200        &1.22$-$1.22\\
             &                 & \textsl{V} &  3,  30,  900        &1.23$-$1.24\\
             &                 & \textsl{I} &  3,  10,  900        &1.19$-$1.19\\
MarkA        & 2010 May 09     & \textit{U} & 240                  &1.30$-$1.30\\
             &                 & \textit{B} & 180                  &1.33$-$1.33\\
             &                 & \textit{V} & 60                   &1.28$-$1.28\\
             &                 & \textit{I} & 60                   &1.26$-$1.26\\
PG~1633      & 2010 May 09     & \textit{U} & 90,120               &1.29$-$1.72\\
             &                 & \textit{B} & 60,90                &1.29$-$1.74\\
             &                 & \textit{V} & 30,40,90             &1.29$-$1.76\\
             &                 & \textit{I} & 30,40                &1.29$-$1.78\\
PG~1323      & 2010 May 09     & \textit{U} & 3x25, 60             &1.07$-$1.88\\
             &                 & \textit{B} & 2x20, 26, 60         &1.07$-$1.71\\
             &                 & \textit{V} & 10, 3x15, 2x30       &1.07$-$1.79\\
             &                 & \textit{I} & 3x15, 30             &1.07$-$1.74\\
NGC~5715     & 2010 May 10     & \textit{U} & 10, 300, 1500        &1.19$-$1.19\\
             &                 & \textit{B} & 10, 200, 1200        &1.25$-$1.26\\
             &                 & \textsl{V} &  3,  30,  900        &1.19$-$1.20\\
             &                 & \textsl{I} &  3,  10,  900        &1.21$-$1.22\\
PG~1323      & 2010 May 10     & \textit{U} & 2x25                 &1.07$-$1.22\\
             &                 & \textit{B} & 2x20                 &1.07$-$1.22\\
             &                 & \textit{V} & 2x10                 &1.07$-$1.20\\
             &                 & \textit{I} & 2x10                 &1.07$-$1.20\\
PG~1657      & 2010 May 10     & \textit{U} & 2x180                &1.29$-$1.49\\
             &                 & \textit{B} & 2x120                &1.29$-$1.52\\
             &                 & \textit{V} & 2x60                 &1.30$-$1.54\\
             &                 & \textit{I} & 2x60                 &1.30$-$1.56\\
\noalign{\smallskip}
\hline
\end{tabular}
\end{table}

\subsection{Photometric reductions}

Basic calibration of the CCD frames was done using IRAF\footnote{IRAF is distributed
by the National Optical Astronomy Observatory, which is operated by the Association
of Universities for Research in Astronomy, Inc., under cooperative agreement with
the National Science Foundation.} package CCDRED. For this purpose, zero exposure
frames and twilight sky flats were taken every night.  Photometry was then performed
using the IRAF DAOPHOT/ALLSTAR and PHOTCAL packages. Instrumental magnitudes were extracted
following the point-spread function (PSF) method (Stetson 1987). A quadratic, spatially
variable, master PSF (PENNY function) was adopted, because of the large field
of view of the two detectors. Aperture corrections were then determined
making aperture photometry of a suitable number (typically 10 to 20) of bright, isolated,
stars in the field. These corrections were found to vary from 0.160 to 0.290 mag, depending
on the filter. The PSF photometry was finally aperture corrected, filter by filter.

\subsection{Photometric calibration}

After removing problematic stars, and stars having only a few observations in Landolt's
(1992) catalog, our photometric solution  for the CTIO run was extracted from
a grand total of 92 measurements per filter, and
turned out to be:\\

\noindent
$ U = u + (3.110\pm0.010) + (0.45\pm0.01) \times X - (0.010\pm0.006) \times (U-B)$ \\
$ B = b + (2.143\pm0.012) + (0.27\pm0.01) \times X - (0.118\pm0.007) \times (B-V)$ \\
$ V = v + (1.740\pm0.007) + (0.15\pm0.01) \times X + (0.038\pm0.007) \times (B-V)$ \\
$ I = i + (2.711\pm0.011) + (0.08\pm0.01) \times X + (0.041\pm0.008) \times (V-I)$ \\

\noindent
The final {\it r.m.s} of the fitting was 0.030, 0.015, 0.010, and 0.010 in $U$, $B$, $V$
and $I$, respectively.\\

\noindent
As for LCO observations, we provided  individual night-based photometric solutions.
However, since the three solutions were identical, we merged and averaged them together in a
single photometric solution. This implies a grand total of 103
measures per filter, and the solutions read:

\noindent
$ U = u + (4.902\pm0.010) + (0.41\pm0.01) \times X + (0.129\pm0.020) \times (U-B)$ \\
$ B = b + (3.186\pm0.012) + (0.31\pm0.01) \times X + (0.057\pm0.008) \times (B-V)$ \\
$ V = v + (3.115\pm0.007) + (0.17\pm0.01) \times X - (0.057\pm0.011) \times (B-V)$ \\
$ I = i + (3.426\pm0.011) + (0.07\pm0.01) \times X + (0.091\pm0.012) \times (V-I)$ \\

The final {\it r.m.s} of the fitting in this case was 0.040, 0.025, 0.015, and 0.015
in $U$, $B$, $V$ and $I$, respectively.\\

\noindent
Global photometric errors were estimated using the scheme developed by Patat \& Carraro
(2001, Appendix A1), which takes into account the errors resulting from the PSF fitting
procedure (i.e., from ALLSTAR/ALLFRAME), and the calibration errors (corresponding to the zero point,
colour terms, and extinction errors). In Fig.~2 we present our global photometric errors
in $V$, $(B-V)$, $(U-B)$, and $(V-I)$ plotted as a function of $V$ magnitude. Quick
inspection shows that stars brighter than $V \approx 20$ mag have errors lower than
$\sim0.05$~mag in magnitude and lower than $\sim0.10$~mag in $(B-V)$ and $(V-I)$. Higher
errors are seen in $(U-B)$.\\

\noindent
Our CTIO final optical photometric catalogs consist of 4432, 8485, 12464, 1163, and
3892 entries having \textit{UBVI} measurements
down to V$\sim$22 mag for ESO131SC09, NGC~5284, VdB-Hagen~164, NGC~6268, and Czernik~38, respectively.
As for LCO, the catalogs for IC~2714, NGC~4052, NGC~5316 and NGC~5715 report 2756,
6650, 2524, and 4105 \textit{UBVI} entries. These catalogs will be made available at the
WEBDA\footnote{http://www.univie.ac.at/webda/navigation.html}
data-base maintained by E. Paunzen at Vienna University, Austria.

\begin{figure}
   \centering
   \includegraphics[width=\columnwidth]{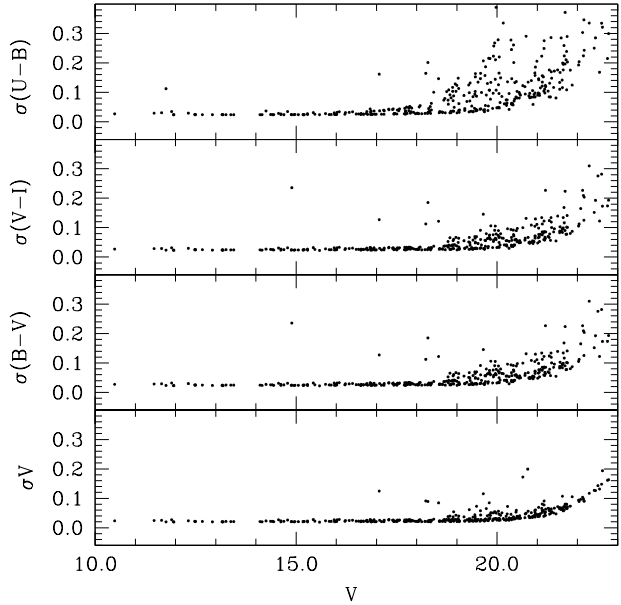}
   \caption{Photometric errors in $V$, $(B-V)$,  $(V-I)$, and $(U-B)$, as a function of the $V$
            magnitude.}
   \end{figure}

\subsection{Completeness and astrometry}

Completeness corrections were determined by running artificial star experiments
on the data. Basically, we created several artificial images by adding artificial stars
to the original frames. These stars were added at random positions, and had the same
colour and luminosity distribution of the true sample. To avoid generating overcrowding,
in each experiment we added up to 25\% of the original number of stars. Depending on
the frame, between 1000 and 5000 stars were added. In this way we have estimated that the
completeness level of our photometry is better than 90\% down to $V = 19.5$.\\

Each optical catalog was then cross-correlated with 2MASS, which resulted in a final catalog
including \textit{UBVI} and \textit{JHK$_{s}$} magnitudes. As a by-product,
pixel (i.e., detector) coordinates
were converted to RA and DEC for J2000.0 equinox, thus providing 2MASS-based astrometry, useful
for {\it e.g.} spectroscopic follow-up.\\

\section{Comparison with previous photometry}
The only cluster in our sample having previous CCD photometry is Czernik~38,
for which BV photometry was provided by Maciejewski (2008).
We can only compare BV, since Maciejewski did not observe in I.
We found 425 stars in common, and the results are shown in Fig~3, in the sense
of our photometry minus  Maciejewski.

\begin{figure}
   \centering
   \includegraphics[width=\columnwidth]{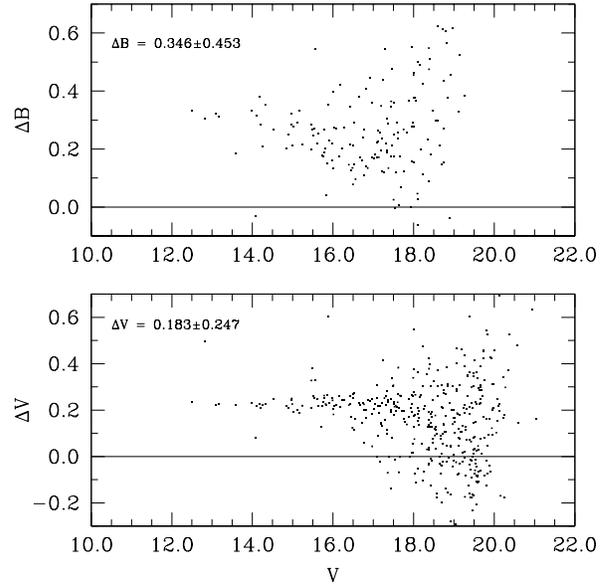}
   \caption{Comparison of our photometry for Czernik~38 with Maciejewski  (2008) for $V$ and $B$.
   The comparison is in the sense  of our photometry versus theirs.}
   \end{figure}

\noindent
We found that the two studies are very different both in V and in B mag.
The mean differences are reported in the top left corners of the various panels in Fig.~3.
The reasons for such important zero point differences can be various. We remind the
reader that Maciejewski observed in very poor seeing conditions ($\sim$ 4 arcsec)
and with a Schmidt camera having a scale as large as 1.08 arcsec/pixel.
Besides, the color range of the standard stars is limited between 0.3 and 1.3 in
B-V, while all Main Sequence stars in Czernik 38 have redder colors. This
implied he had to extrapolate colors when calibrating. Besides, with such a
poor detector scale, the seeing conditions
and moderated crowding of the field (see Fig.~1)  easily produce blends, and stars tend, on
the average, to be brighter.
This is exactly what the positive residuals in Fig.~3 indicate.
For all these reasons we believe our photometry is more solid and precise (note the scatter in the residual
in Fig~3 for V larger than $\sim$ 17.0 mag).\\

\noindent
To further check the quality of our data, we compared our UBV photometry for IC~2714 
with the higher quality photoelectric study by Clari\'a et al. (1994).
This is shown in Fig.~4.
The comparison is done for 90 common stars and it is in the sense of our photometry minus
Clari\'a et al. (1994).
The results are quite good for V and B-V, as indicated in Fig.~4, down to V $\sim$ 14.0.
However, we found some discrepancy
for U-B, in the form of a un-accounted color term.

\begin{figure}
   \centering
   \includegraphics[width=\columnwidth]{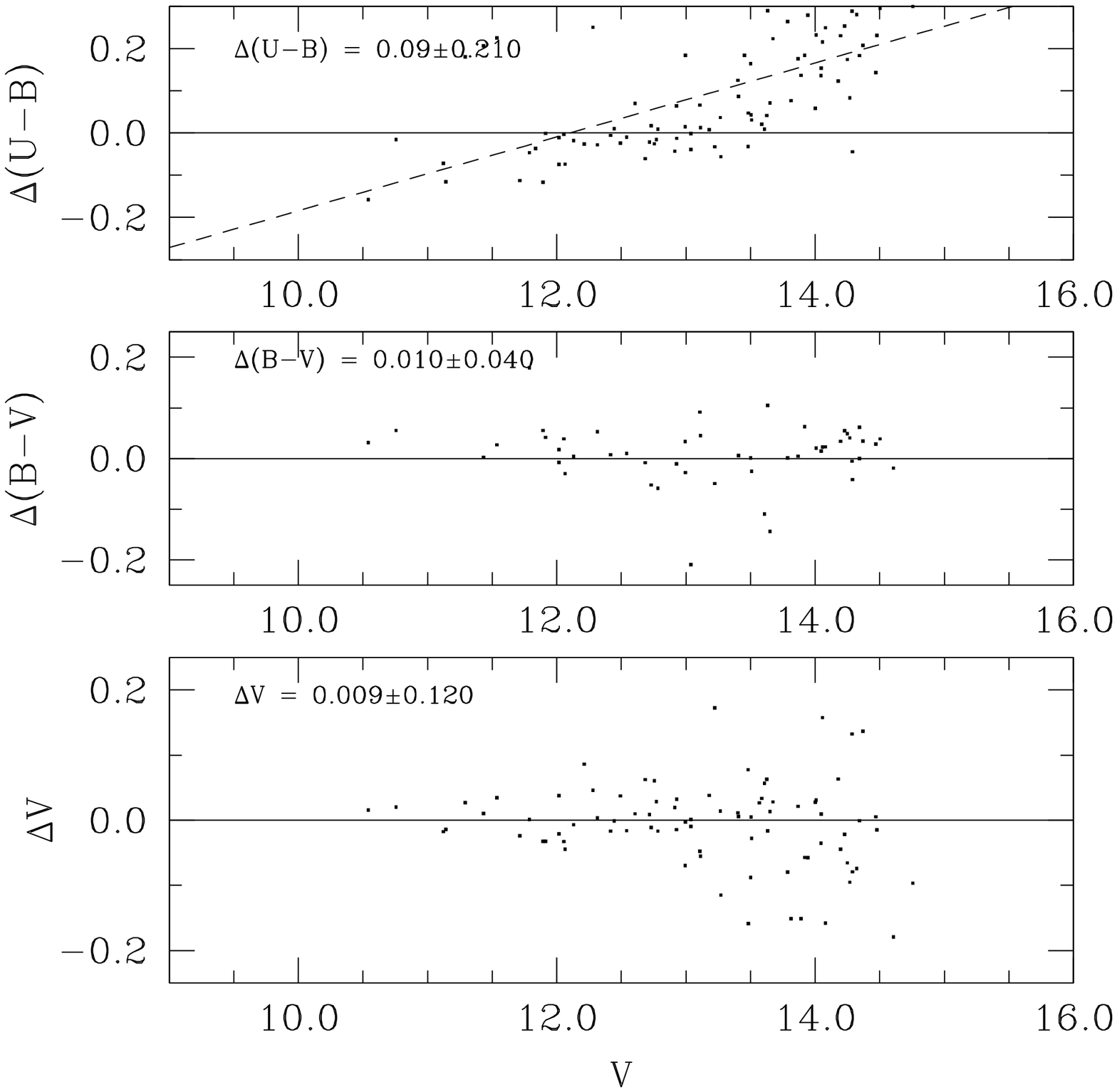}
   \caption{Comparison of our photometry for IC~2714 with Clari\'a et al. 1994 for V, B-V, and U-B.
   The comparison is in the sense  of our photometry versus theirs.}
   \end{figure}

\begin{figure*}
   \centering
   \includegraphics[width=15truecm]{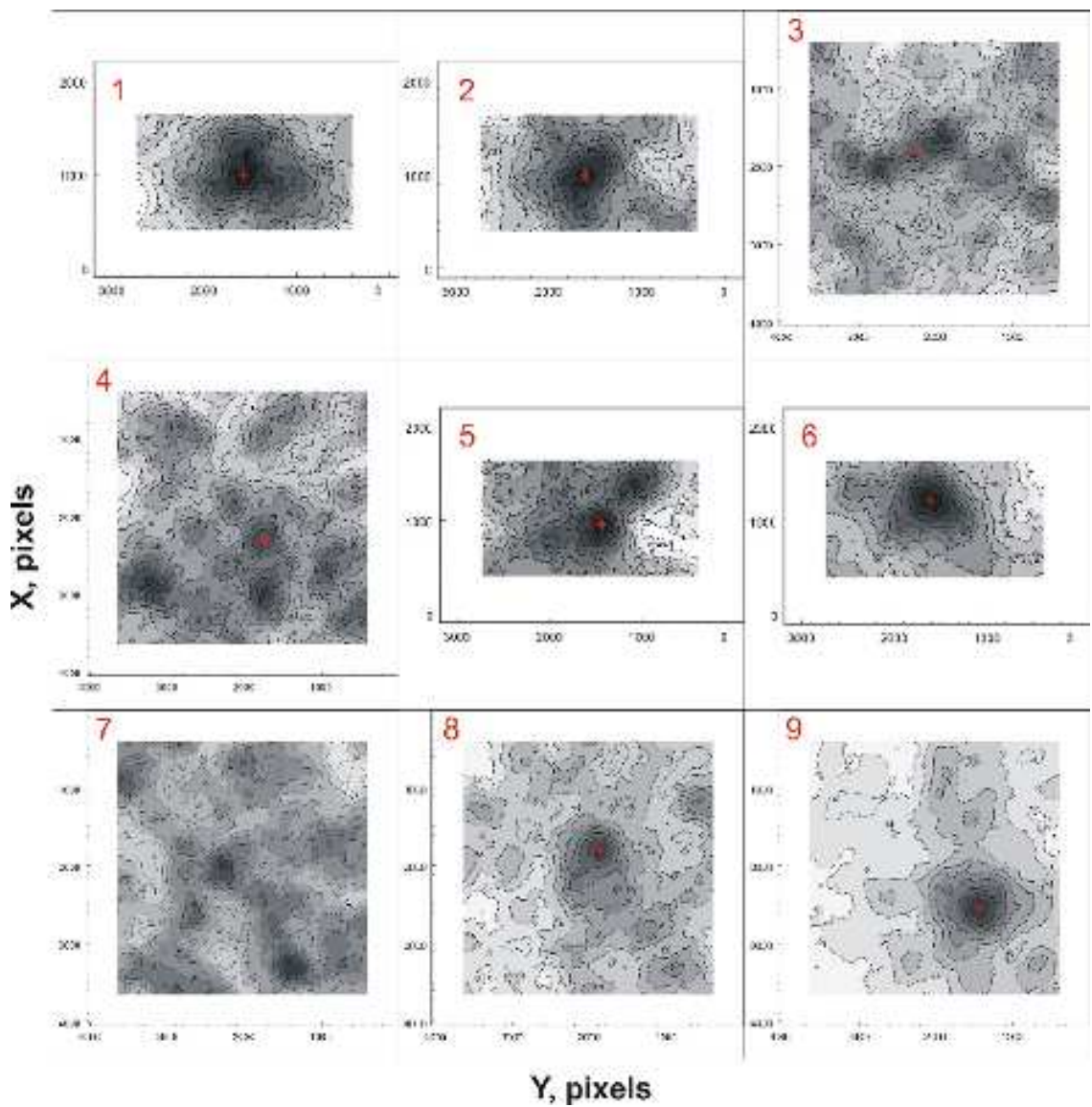}
   \caption{Optical surface density maps for the nine program clusters. North is right, East is down.
   Numbering follows Table~1.}
   \end{figure*}

\begin{figure*}
   \centering
   \includegraphics[width=15truecm]{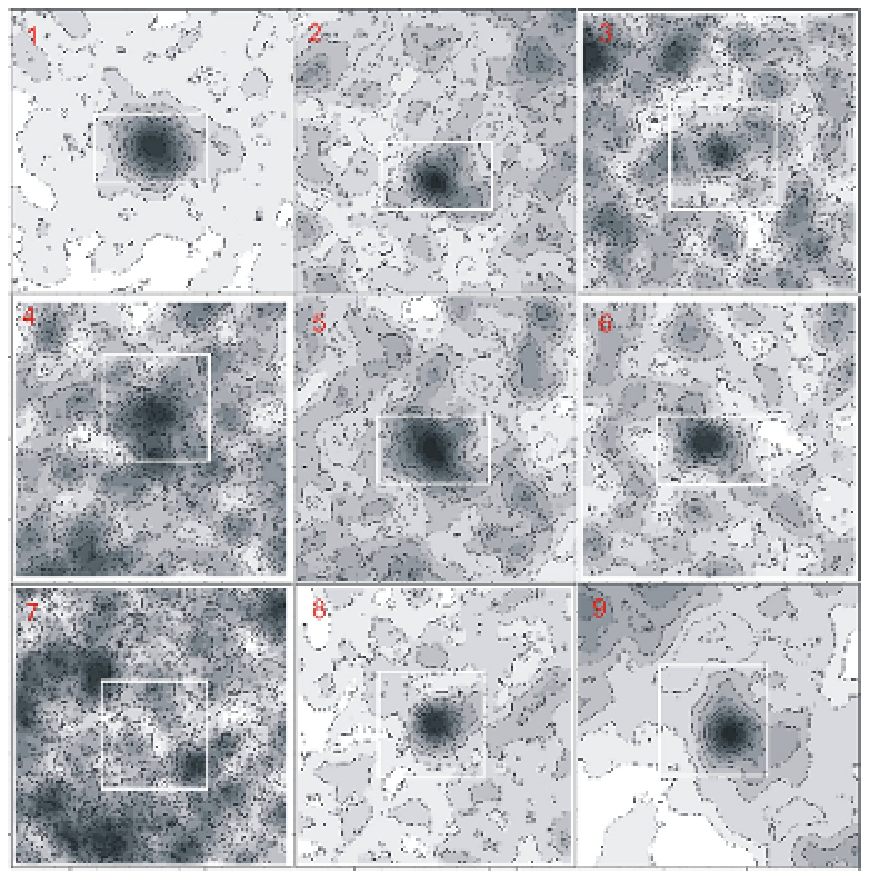}
   \caption{2MASS Surface density maps for the nine program clusters. North is right, East is down.
   White boxes enclose the area covered by optical data. Numbering follows Table~1.}
   \end{figure*}

\section{Star counts, clusters' reality and size.}
In this section we make use of our photometric data-set, and of infra-red
photometry from the 2MASS data-base, to perform star counts in the area of the program
clusters, assess their reality, and derive estimates of their radii.\\

\subsection{Surface Density Maps and Cluster Center Coordinates}

Surface Density Maps (SDM) were constructed
for all fields under investigation to determine the cluster's
reality and size. Examples of the application of this technique
can be found in
Prisinzano et al. (2001), Pancino et al.(2003), and Seleznev et al (2010),
which the reader is referred to for any additional details.\\

Briefly, SDM were constructed using the kernel estimation method (see {\it e.g.}
Silverman 1986), with a kernel half-width of 400 pixel s(corresponding
to 2.90$\arcmin$ for LCO clusters and 1.93$\arcmin$ for CTIO clusters),
Moreover,  a grid of 20-pixel cells for LCO clusters, and of
50-pixel cells for CTIO clusters, were adopted.  The large kernel half-width
(HW) was chosen in order to both decrease the effect of density fluctuations
(to avoid, for example, numerous un-real density peaks inside a cluster),
and to detect the cluster centre more clearly. Only stars brighter
than a specific  magnitude limit -depending on the field- were considered since
the inclusion of faint stars has sometimes the effect of
confusing the cluster inside the rich surrounding back/fore-ground. These limiting
magnitudes are listed in Table~4.\\

\begin{table}
\tabcolsep 0.1truecm
\caption{Limiting magnitude for star counts in optical.}
\begin{tabular} {|r|r|r|}
\hline
label &  Name  &  limit in V\\
\hline
1 & IC 2714 & 16 \\
\hline
2 & NGC 4052 & 16 \\
\hline
3 & ESO131SC09 & 17 \\
\hline
4 & NGC 5284 & 17 \\
\hline
5 & NGC 5316 & 16 \\
\hline
6 & NGC 5715 & 16 \\
\hline
7 & VdB-Hagen 164 & 17 \\
\hline
8 & NGC 6268 & 17 \\
\hline
9 & Czernik 38 & 19 \\
\hline
\end{tabular}
\end{table}

\noindent
To avoid under-sampling, we only made
use of the 1240 $\times$ 2340 pixel ($\sim9.0\arcmin\times17.0\arcmin$)
central region for LCO clusters  and
of the 3250 $\times$ 3250 pixel ($\sim15.7\arcmin\times15.7\arcmin$)
central region for CTIO clusters.
The resulting SDMs are shown in Fig.~5, where the iso-density contour
lines are in units of (100 pixel)$^{-2}$.\\

\noindent
New rough coordinates for the clusters' centres were  obtained from the
centre of symmetry of the inner (maximum) density contours.
In the cases of irregular structures we estimated centre position
through comparison with 2MASS density maps (see discussion below).
The new cluster centres coordinates are listed in
Table~5, and they are indicated by crosses in Fig.~5. We emphasize 
that the use of more sophisticated methods for cluster centre determination
would not make much sense in this case, because the position of the cluster
centre clearly depends both on limiting magnitude and on kernel half-width.
Table~5 also contains  rough estimates of radii for the  denser central
regions of the clusters ({\it cores}).\\

\begin{table}
\tabcolsep 0.1truecm
\caption{ Center coordinates and core radii in pixels of the nine clusters
under study.}
\begin{tabular} {rrcccr}
\hline
label &  Name  &  $X_c$  &  $Y_c$  &  "core" radius &  comments\\
\hline
  &  &  pixels  &   pixels  &  pixels  &  \\
\hline
1 & IC 2714 & 1000 & 1570 & 1000 &  \\
\hline
2 & NGC 4052 & 1000 & 1600 & 800 &  \\
\hline
3 & ESO131SC09 & 1860: & 2280: & 900 & two centers\\
\hline
4 & NGC 5284 & 2300:  & 1720: & 1050 & complicated structure  \\
\hline
5 & NGC 5316 & 950 & 1450 & 800 &  \\
\hline
6 & NGC 5715 & 1200 & 1600 & 600 & \\
\hline
7 & VdB-Hagen 164 & - & - & - &  \\
\hline
8 & NGC 6268 & 1800 & 1830 & 600 & \\
\hline
9 & Czernik 38 & 2550 &  1450 &  700 & \\
\hline
\end{tabular}
\end{table}

\noindent
One can clearly notice that in most cases the cluster size is larger than
the detector field of view  or just comparable to it. 
Therefore, in order to study the cluster structure
in larger areas, to plot density profiles, and to  estimate cluster sizes,
we made use of  star counts of photometry from the 2MASS data-base.\\

\subsection{2MASS Surface Density Maps and Radial Surface Density Profiles for
Sample Clusters}

Nine fields 3600x3600 arcsec (boxes) centered on clusters
were downloaded from 2MASS data-base. These fields contain huge amounts of  stars
and SDM plotted taking into account all stars to show density fluctuations
which mask the clusters completely. In order to make clusters more evident, SDM were plotted for
stars selected on color-magnitude diagrams (CMD). Stars were selected
into intervals (J-H)$\simeq\pm$0.15 wide around the position of isochrones' set
super-imposed on the CMD and using preliminary values of reddening and distance
modulus, mostly based on visual inspection.\\

\noindent
SDM's obtained with kernel half-width of 5 arcmin and a grid of 0.5-arcmin cells
are shown in Fig.~6. Limiting magnitudes of stars were selected in order to
make clusters as visible as possible, and these limiting magnitudes are listed
in the Table~6 (third column).\\

\begin{table}
\tabcolsep 0.1truecm
\caption{Limiting magnitudes for density maps from 2MASS.}
\begin{tabular}{rrrr}
\hline
label &  Name  &  SDM limit in J & RSDP limit in J\\
\hline
1 & IC 2714 & 13   & 15  \\
\hline
2 & NGC 4052 & 13   & 16  \\
\hline
3 & ESO 131 & 12   & 14  \\
\hline
4 & NGC 5284 & 13   & 15  \\
\hline
5 & NGC 5316 & 13   & 15  \\
\hline
6 & NGC 5715 & 13   & 16  \\
\hline
7 & BH 164  & 13   &      \\
\hline
8 & NGC 6268 & 13   & 16  \\
\hline
9 & Cz 38   & 15   & 16  \\
\hline
\end{tabular}
\end{table}

To guide the reader, white rectangles in  Fig.~6 shows position 
and extent of optical fields of view.
These positions were obtained by cross-identification of the brightest stars
in the optical and infrared fields. All clusters clearly stand out
except for VdB-Hagen 164. This cluster from the list of van den Bergh
and Hagen (1975), if real,  is probably very poorly populated, close,
very sparse with an angular size of supposedly much more than one  degree.
The data we have analyzed, however, do not give us the impression
of a cluster, and therefore we are going to consider it as a false detection
in this paper (see the following Sections).\\

\noindent
SDM for all the clusters were obtained for different  limiting magnitudes.
A close analysis of these maps shows that the cluster center position depends
on limiting magnitude. Table~7 shows coordinates of 2MASS fields
and cluster center positions with respect to field center (telescope pointing):
$\alpha_c=\alpha+\Delta\alpha$, $\delta_c=\delta+\Delta\delta$ (corrections
are given in arcminutes). Each limiting magnitude corrections are
given in the order of $\Delta\alpha/\Delta\delta$.

\begin{table*}
\tabcolsep 0.1truecm
\caption{New equatorial coordinates for the clusters under investigation,
obtained using 2MASS-based surface density maps.}
\begin{tabular} {rrrrrrrrrr}
\hline
label &  Name  &  RA & Dec & $J<12 $ & $J<13 $ & $J<14 $ &
 $J<15 $ & $J<16 $ & Comments \\
\hline
1 & IC 2714 & $11^h17^m30^s$ & $-62^{o}44^{\prime}00^{\prime\prime}$ &
   & -0.6/+0.5 & -1.0/+0.3 & -0.5/+1.0 &  &  \\
\hline
3 & ESO 131 & $12^h29^m38^s$ & $-57^{o}52^{\prime}36^{\prime\prime}$ &
 +1.0/+0.5 & +1.0/+0.6 & +0.1/+0.6 &   &   &  \\
\hline
2 & NGC 4052 & $12^h01^m12^s$ & $-63^{o}13^{\prime}00^{\prime\prime}$ &
   & +5.8/+0.1 & +5.0/-0.1 & +4.5/-0.1 & +4.0/-0.5 &  \\
\hline
4 & NGC 5284 & $13^h47^m23.3^s$ & $-59^{o}08^{\prime}58^{\prime\prime}$ &
   & -4.2/+1.1 & -3.8/+0.8 & -2.4/+0.5 & -1.3/-2.3* & *2nd center became main \\
\hline
5 & NGC 5316 & $13^h53^m54^s$ & $-61^{o}52^{\prime}00^{\prime\prime}$ &
   & +2.0/-0.7 & +3.0/-0.5 & +4.4/-0.1 &  &  \\
\hline
6 & NGC 5715 & $14^h43^m24^s$ & $-57^{o}33^{\prime}00^{\prime\prime}$ &
   & +0.9/-2.5 & +1.0/-2.5 & +1.0/-2.5 & 0.0/-2.2 &  \\
\hline
7 & BH 164 & $14^h48^m14^s$ & -$66^{o}20^{\prime}23^{\prime\prime}$ &
 -  & - & - & - & - &  \\
\hline
8 & NGC 6268 & $17^h02^m10^s$ & $-39^{o}43^{\prime}42^{\prime\prime}$ &
   & -0.3/-0.1 & -0.2/+0.2 & -0.1/+0.3 & +0.2/0.0 &  \\
\hline
9 & Cz 38  & $18^h49^m42^s$ & $ 04^{o}56^{\prime}00^{\prime\prime}$ &
   & +2.4/+2.1 & +1.5/+2.0 & +1.5/+2.0 & +1.1/+1.8 &  \\
\hline
\end{tabular}
\end{table*}

Radial Surface Density Profiles (RSDP) - F(r) - were then obtained by using
the kernel method, and adopting
a kernel half-width of 2 arcminutes. The kernel size was adjusted 
to avoid both extra smoothing and extra irregularities.
RSDP are showed in Fig.~7 as solid curves with points. Solid curves
show 1-$\sigma$ confidence interval and is obtained using a "smoothed bootstrap"
algorithm (see {\it e.g.} Seleznev et al. (2000) and Prisinzano et al. (2001) for additional
details).
Solid polygonal lines shows density hystograms with steps of one arcmin.
The adopted limiting magnitudes for RSDP are listed in the fourth column of
Table~6. Finally, F(r) is shown in units of (arcmin)$^{-2}$\\

\noindent
By inspecting Fig.~7 in details, one can notice the small values of
F(r) at the cluster centres' position in a few cases.
These are due to irregularities in the field caused, in some cases,
by patchy extinction and/or field density fluctuations.
Technically, the reason is that the cluster centres were determined
from SDMs constructed with a large kernel half-width (5 arcmin),
whereas the RSDPs have been derived adopting smaller values for the kernel
width. The smaller scale produces a fluctuating profile, and as a consequence,
low density values can be obtained at the centres
when F(r) is constructed  using small increments.\\

\begin{figure}
   \centering
   \includegraphics[width=\columnwidth]{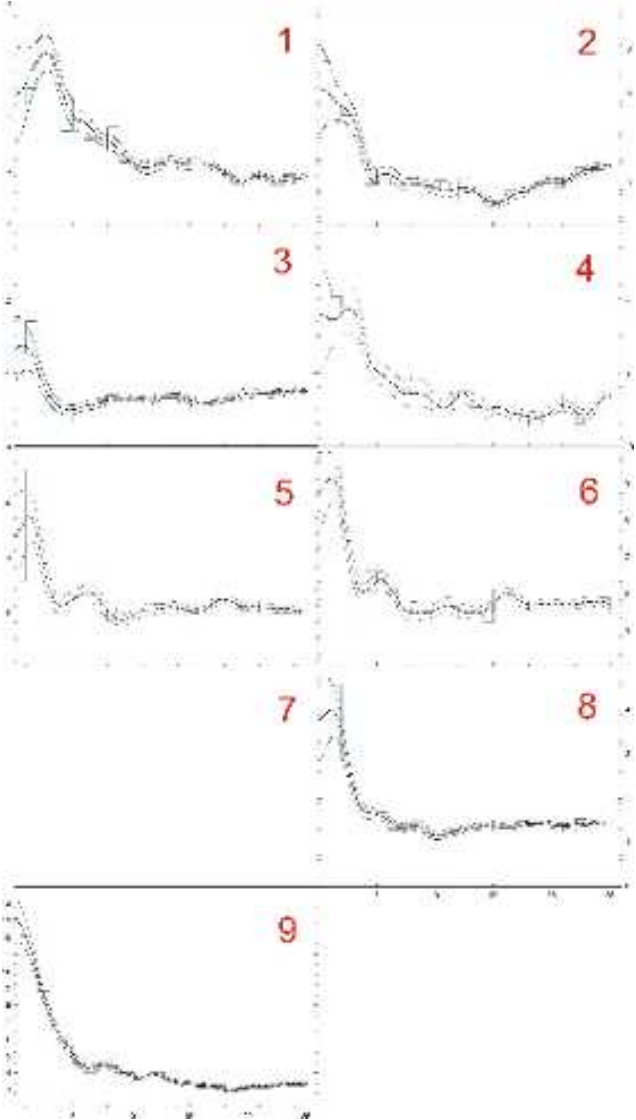}
   \caption{Surface radial density profiles for the nine program cluster. Numbering follows Table~1.
   No profile is shown for VdB-Hagen~164, see Section 5.}
   \end{figure}

\subsection{Results from the SDM and RSDP analysis}
In this section we analyse the outcome of star counts on a cluster-by-cluster basis,
highlighting the most important results.\\

\noindent
{\bf 1. IC 2714}\\
This is a rich cluster with an angular radius of about 19 arcminutes. The
cluster manifests a
relatively regular structure for $J_{lim}=13$, which became more irregular
for $J_{lim}=15$.  However, at $J_{lim}=16$ (the limit of 2MASS  completeness)
the cluster disappears against background density fluctuations.\\

\noindent
{\bf 2. NGC 4052}\\
This cluster has an angular radius of about 14 arcminutes. It is well
seen in the SDMs down to $J_{lim}=14$. When including fainter stars, the cluster
is still visible, but looks less populous than neighbour density fluctuations.
"Optical" SDM shows a double core and an elongated outer core region.\\

\noindent
{\bf 3. ESO 131SC09}\\
This is small group with angular radius of about 4 arcminutes. SDM shows
two neighbour density fluctuations, 9 arcmin to the South, and about 8 arcmin
to the North-West, respectively. 
To address the question whether or not they are part to the cluster, one needs
additional information, {\it e.g.} proper motions. The {\it optical } SDM shows
elongation of cluster roughly in the North-South direction and the presence of
several density peaks. The position of cluster center was conservatively taken as 
the small peak just in between the two more general density peaks. In the
2MASS SDM the cluster is visible down to $J_{lim}=14$ 
and disappears when including fainter stars against the background density fluctuations.\\
Interestingly enough, the star density peak has a very low contrast with the field,
the lowest in our sample. This fact may indicate that the central peak can be caused by
an inter-stellar absorption minimum.\\

\noindent
{\bf 4. NGC 5284}\\
This cluster has a complex structure. At $J_{lim}=13$ and $J_{lim}=14$ SDMs
show a double structure with the second component 8 arcmin to the East.
At $J_{lim}=15$ this second component disappears, but we see another one
13 arcmin apart, in the Noth-East direction. At $J_{lim}=16$ this component becomes
prominent. The {\it optical}  SDM also shows a complex structure with several
secondary maxima. The RSDP gives an estimate of cluster radius of about
17 arcminutes.\\

\noindent
{\bf 5. NGC 5316}\\
This cluster stands out neatly  in  SDM for $J_{lim}=13$ and $J_{lim}=14$.
At $J_{lim}=15$ cluster is still visible, but we see comparable density
fluctuations around them. At $J_{lim}=16$ the cluster disappears against
the background density fluctuations. The RSDP suggests a  cluster
radius of about 9 arcminutes. Optical SDM shows asymmetric complex
structure with at least two secondary density maxima. \\

\noindent
{\bf 6. NGC 5715}\\
This cluster is well defined for all limiting magnitudes. At $J_{lim}=14$
it exhibits a  double structure. Neighbour density fluctuations become
stronger when including fainter stars. RSDP gives a  cluster
radius estimate of about 8 arcminutes.\\

\noindent
{\bf 7. VdB-Hagen 164}\\
This cluster was listed in the paper of van den Bergh and Hagen (1975).
It does not possess a clear density peak either in "optical" or in "infrared" SDM.
van den Bergh and Hagen (1975) write that cluster is visible in blue
plates and not visible in red plates. Possibly it is a sparse group of young
stars with an angular size of about a degree or even more. However, SDM plotted for
a field 2x2 degrees large, do not show any density peak either. \\

\noindent
{\bf 8. NGC 6268}\\
This cluster is clearly defined for all limiting magnitudes. Density
fluctuations grow with increasing  limiting magnitude and
concentrate in the North-West quadrant with respect to the cluster.
RSDP gives a cluster radius estimate of about 10 arcminutes. \\

\noindent
{\bf 9. Czernik 38}\\
This is a relatively rich cluster with an angular radius of about 16-18 arcminutes.
It has a symmetric core and a slightly asymmetric halo elongated in the North
direction. RSDP of this cluster shows a "step" near r=8 arcmin.\\

\begin{figure*}
   \centering
   \includegraphics[width=15truecm]{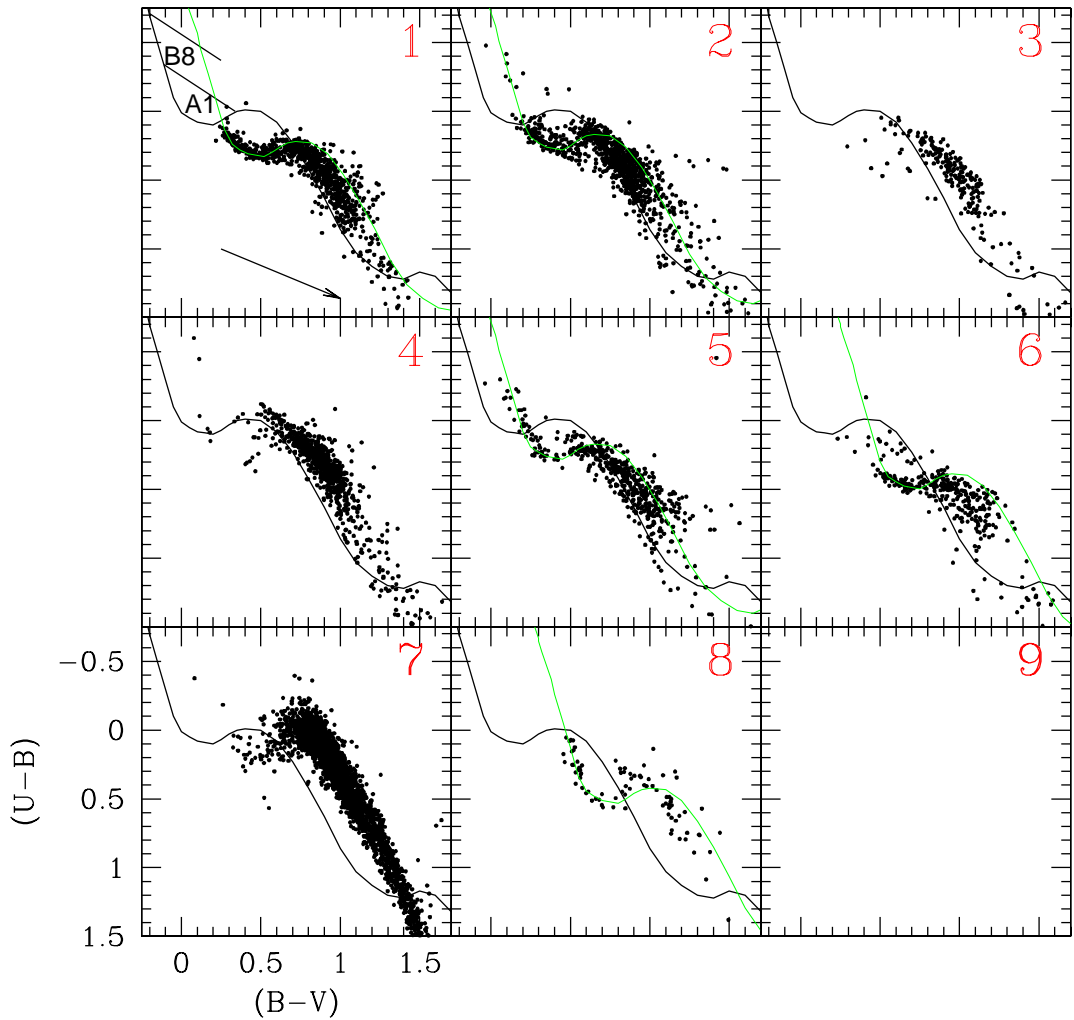}
   \caption{Color-color diagrams for the nine program clusters. The numbering follows
   Table~1. The solid line is an empirical zero reddening Zero Age Main sequence.
   The arrow in the top-left panel indicates the reddening vector. Two spectral type are
   indicated to guide the eye, together with the reddening path. The green line is
   the same ZAMS, but shifted along the reddening path by an amount corresponding to
   the cluster reddening E(B-V).}
   \end{figure*}

\section{Fundamental parameters: reddening, distance and age}
In this Section we make use of the results of the star count analysis
to derive estimates of the fundamental parameter, namely reddening, distance,
and age for the clusters under study.

\subsection{Basics of the method}
We only considered the stars that fall inside the core radius,
as defined in previous Sections, with the aim to alleviate as much as
possible  field star contamination and render the cluster more visible.
The method we employed is based first on the inspection of the
color-color diagram (TCD), in the B-V vs U-B color combination, 
to derive an independent estimate of the cluster mean reddening.
In this diagram, the position of stars with spectral types
earlier than A0, only depends on reddening (Carraro et al. 2008).
Then, the analysis moves to the inspection of the color magnitude
diagrams (CMD),
in various color combinations, to derive estimates of the cluster distance and age.
For the sake of homogeneity with previous studies ({\it e.g.} Seleznev et al. 2010)
we adopt here $R_{\odot}$=8.5 kpc
as the distance of the Sun to the Galactic center, and $R_{V}$ = 3.1 as the ratio
of total to selective absorpion $\frac{A_V}{E(B-V)}$.
We stress, finally, that we are going to adopt solar metallicity for these
clusters, as a working hypothesis, when no information is  available from
spectroscopy. This is partly justified by the clusters
location in the inner disk. Should solar metallicity clearly be unsuitable,
we will then explore different values.

\subsection{Error analysis}
With  the probable exception of IC~2714, most clusters in this paper are being
studied for the first time. 
We are, therefore, facing 
the  well-known problem of associating reliable errors
to distance and age. Without precise estimates of reddening
and metallicity, it is extremely difficult to perform a proper
error assessment. This would imply, in theory, a full error propagation which
would in general produce
a very  large iper-volume in the parameters' space with many
solutions which would not pass a  simple by-eye inspection.\\

\noindent
We will, therefore, limit ourselves to provide fitting errors
for the cluster reddening and apparent distance moduli, being totally aware
that they most probably are only rough lower limits awaiting improvements as
soon as more precise metallicity measurements will be available.
However, in deriving distance, a full propogation is done
taking into account the whole range of values for reddening and distance modulus.
Finally, as far as the ages is considered, only fitting errors are reported, adopting
solar metallicity (see below).

\subsection{Clusters' reddening}
In Figure 8 we show the TCDs for eight of the nine program clusters.
Unfortunately, we could not provide U-band photometry for Czernik~38 and therefore
we are going to estimate its reddening simultaneously with age and distance
from the CMD analysis, using theoretical isochrones, in a less effective way (see below).
In each of the panel in Fig.~8 we show the TCD for the program
clusters following their numbering as in Table~1. As anticipated, only
stars within the core radius are used.
The solid line is  a zero reddening, solar metallicity,  
empirical zero age main sequence (ZAMS) taken
from Schmidt-Kaler(1982).
In each panel, the same empirical ZAMS is shifted along the reddening vector
(indicated by the arrow in the upper-left corner)
to fit the bulk of the stars in each cluster, using the green color, whenever
this is possible.
As for the fit, this is done by considering only the stars with spectral type
earlier than A0, since for later spectral type stars there is not a unique
fitting solution. To facilitate viewing, we indicated two spectral types along the
ZAMS and the paths, parallel to the reddening vector, along which stars are
displaced by reddening.

This procedure allows us to estimate the mean reddening for each cluster, that we report
in Table~8 (third column). The associated uncertainties have been estimated visually 
and represent the range in reddening we can move back and forth the ZAMS
keeping the fit acceptable. This uncertainty  basically comes from photometric errors
and variable reddening across the cluster. When the uncertainty is larger than
photometric errors, which are much less than 0.03 mag in this magnitude range (see Fig~2),
we are forced to  conclude that differential reddening is present in the cluster.
This is far from being un-expected, since these clusters are typically located at low
latitude in the inner Galactic disk, inside or close to gas- and dust-rich spiral
features.

\subsection{Clusters' distance and age}
Distances are estimated in the CMDs in Figs.~9 to 11,
using the same ZAMS as in the previous Section, and adopting
the reddening values already derived.
In this process, therefore, distance
is the only free parameter, having fixed metallicity and reddening.
In the same way as for reddening, the uncertainty in distance is estimated
vertically shifting the ZAMS (green line) for the range of distance moduli which provides
an acceptable fit. In this process we pay attention that the reddening 
remains within the range of values independently
estimated in the TCD. \\

The final step is to estimate the age, and for this we make use of isochrones
from the Padova suite of models (Marigo et al. 2008, red symbols).
When fitting isochrones, we pay attention to the turn off (TO) color, magnitude and
shape and, when it is present, to the Red Giant clump color and magnitude.
In addition, we require that the fit is of the same quality in all the three color
combination CMDs.\\
The results are summarized in Table~8, and illustrated in Figs.~9 to 11.

\begin{figure*}
   \centering
   \includegraphics[width=15truecm]{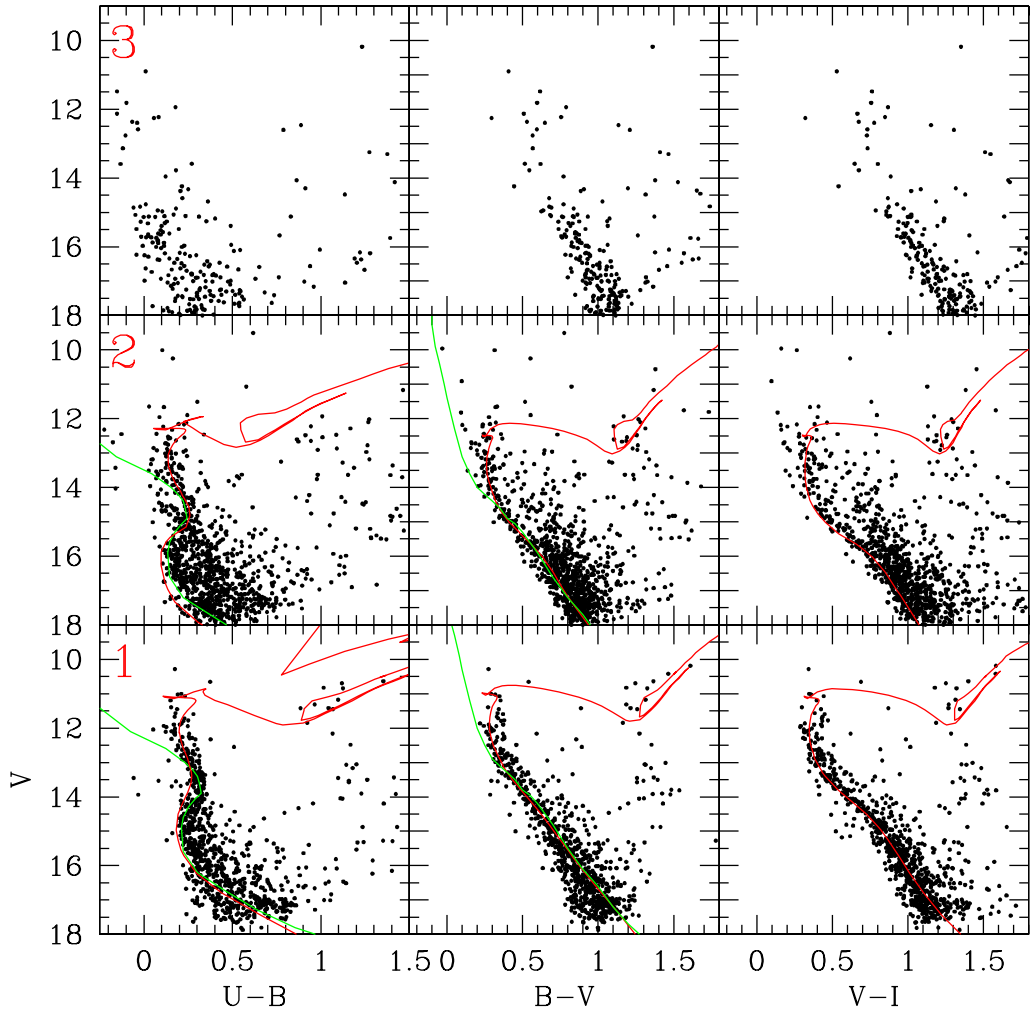}
   \caption{Color Magnitude Diagrams in the V/U-B, V/B-V, and V/V-I for
    IC~2714, NGC~4052, and ESO131SC09. Numbering follows Table~1. The green line
    is an empirical ZAMS shifted by an amount corresponding to the apparent distance modulus
    to fit the cluster sequence. The red lines are isochrones from Marigo et. al. 2008}
   \end{figure*}

\begin{figure*}
   \centering
   \includegraphics[width=15truecm]{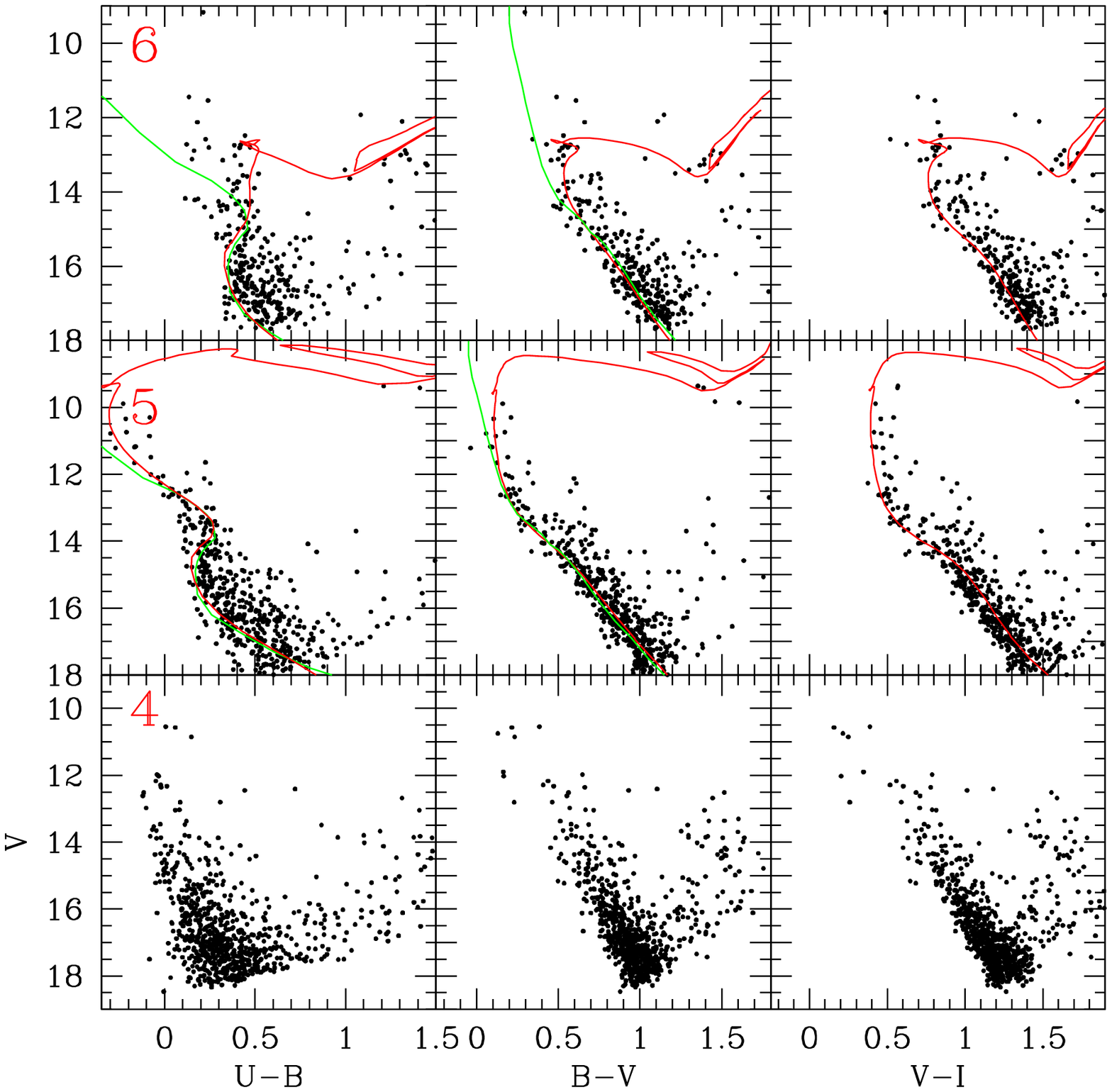}
   \caption{Color Magnitude Diagrams in the V/U-B, V/B-V, and V/V-I for
    NGC~5284, NGC~5316, and NGC~5715. Numbering follows Table~1. The green line
    is an empirical ZAMS shifted by an amount corresponding to the apparent distance modulus
    to fit the cluster sequence. The red lines are isochrones from Marigo et. al. 2008.}
   \end{figure*}

\begin{figure*}
   \centering
   \includegraphics[width=15truecm]{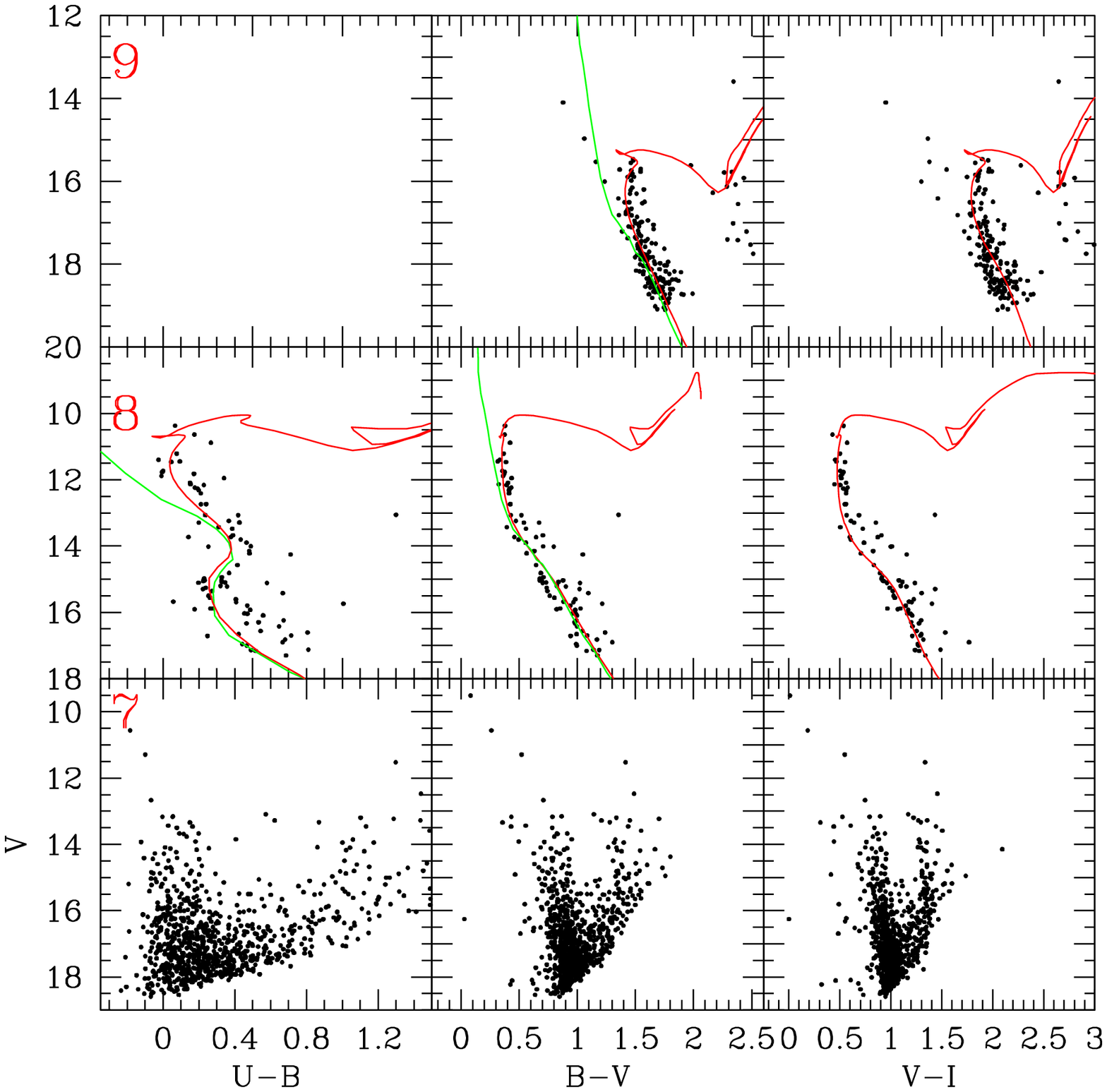}
   \caption{Color Magnitude Diagrams in the V/U-B, V/B-V, and V/V-I for
    VdB-Hagen~164, NGC~6268, and Czernik~38. Numbering follows Table~1. The green line
    is an empirical ZAMS shifted by an amount corresponding to the apparent distance modulus
    to fit the cluster sequence. The red lines are isochrones from Marigo et. al. 2008.}
   \end{figure*}

\section{Discussion}
Having detailed how fundamental parameter estimates have been searched
for, we now discuss  each individual cluster, commenting on the specific results. In fact,
while in general a homogeneous technique has been applied, some cases,
like {\it e.g.} Czernik~38, do require  more information.\\

\noindent
{\bf IC 2714:}\\
The set of fundamental parameters we found for this cluster is in fine 
agreement with Clari\'a et al.(1994)
within the uncertainties, in spite of the color term problem we found when comparing their
photoelectric photometry with our CCD data-set.  
Our results are shown in the bottom panels of Fig.~9.
The TO point is located at V $\sim$ 12.5, (B-V) $\sim$ 0.30,
and (V-I) $\sim$ 0.40. A sparse red giant clump is visible at V$\leq$ 12.0, (B-V) $\geq$1.2.
Santos et al. (2009) derived spectroscopic iron abundance of three giants, and provided
for IC~2714 [Fe/H]=0.02$\pm$0.01 and [Fe/H]=-0.01$\pm$0.01, depending
on the line list used. In any case metallicty is very close to solar.

We, therefore, searched the solar metallicity isochrone which best fits the cluster stars
distribution in the CMD, and obtained an age of 3.0$\pm$0.2 x10$^{8}$ years.
This implies an apparent distance modulus (m-M)$_V$=11.6$\pm$0.1.
At the corresponding  helio-centric distance of 1.3$^{+0.10}_{-0.15}$ kpc 
and $\sim$ 40 pc below the Galactic plane,
IC~2714 is most probably on the act of leaving the Carina arm, 
where it formed. 
To assess this scenario, we used available literature 
to derive Galactic heliocentric velocities for IC~2714. The cluster has a radial
velocity $R_{V}$= -13.54$\pm$0.50 km/sec (Mermilliod et al 2008), and UCAC2 proper motions
$\mu_{\alpha} \dot  cos\delta$= -7.55$\pm$0.10 and $\mu_{\delta} = 0.49\pm0.10$
(Dias et al. 2006). 
From these values we derived U = 34.43$\pm$3.18 km/sec, V = 262.51$\pm$6.09  km/sec
and W = 14.41$\pm$1.53 km/sec, where U is positive toward the
anticenter direction, V is positive in the Galactic rotation
sense, and W points toward the North Galactic Pole.
The velocities we derive imply that the cluster is moving apart from the Carina
arm, drifting toward larger Galacto-centric distances.\\

\noindent
{\bf NGC 4052:}\\
The middle panels in Fig~9 show our results for NGC~4052, and the estimates of its
basic parameters are listed in Table~8.
The cluster TO is located at V $\sim$ 13.5, (B-V) $\sim$ 0.30,
and (V-I) $\sim$ 0.40, while a sparse red clump is at V $\sim$12.5, B-V $\sim$1.2
We fit the cluster sequence with a solar metallicity
400 Myr isochrone, and with the Schmidt-Kaler(1982) ZAMS, which both yield
an apparent distance modulus (m-M)$_V$=12.7$\pm$0.2.\\
As a consequence, NGC~4052 is ubicated at 2.2$^{+0.50}_{-0.30}$ kpc from the Sun, and at 7.8 kpc
from the Galactic center.\\

\noindent
{\bf ESO131SC09:}\\
Star counts reveal the presence of a concentration of stars at the position of this object.
However, from the inspection of photometric diagrams, we are keen to derive a different
conclusion (see Fig~9, upper panels).
Basically, we recognize two distinct groups of stars. A small number of bright
blue stars in the upper region of the CMD is in fact super-imposed to a larger group of F-G stars most probably
belonging to the general Galactic disk field. The two groups are clearly separated in the CMDs,
and such a significant magnitude gap excludes that the two groups belong to the same system.
The apparent concentration in the maps is produced by the small group of bright stars
which, evidently, do not define any stellar cluster, but, maybe,  an open cluster remnant
following the Loden (1973) original definition (see also Platais et al. 1998 and Carraro 2006).
For these reasons we refrain from trying any isochrone fitting.\\

\noindent
{\bf NGC 5284:}\\
As star counts showed, this is a star agglomerate with quite a complicated structure. On DSS
or CCD images,
the clusters does not appear at all, and the CMDs shown in the bottom panels
of Fig~10 do not show convincingly the presence of a star cluster. We believe the over-density
is produced by a random concentration of a few blue stars, brighter than V = 13.
As in the case of the previous cluster, this group looks like an open cluster remnant.
In most cases, these groups do not turn out to be real clusters when
proper motions or radial velocities are available to assess individual star
membership to a cluster, as demonstrated by {\it e.g.} Odenkirchen \& Soubiran (2002), Villanova et al. (2005),
Carraro et al. (2005a), and Moni Bidin  et al. (2010).
Therefore, as in the case of ESO131SC09, we refrain from  trying any isochrone fitting.\\

\noindent
{\bf NGC 5316:}\\
The CMDs for this cluster are shown in the middle panels of Fig.~10. The cluster is confirmed to be
a physical group of young stars, with stars having spectral types as 
early as B5-B7. We use the reddening estimated in Fig.~8
and fit the CMDs with solar metallicity isoschrones. The best fit is achieved with an age of 100
Myrs, which, in turn, provides an apparent distance modulus (m-M)$_V$ = 11.50$\pm$0.20.
Therefore,  we place NGC~5316 at an helio-centric distance of 1.4$^{+0.15}_{-0.20}$ kpc and at a Galactic
center distance of 7.6 kpc. 
This cluster lies very close to the formal Galactic plane, at Galactic latitude b = 0$^{o}$.\\

\noindent
{\bf NGC 5715:}\\
The CMDs of NGC~5715 (upper panels of Fig~10) clearly indicate it is an intermediate-age cluster,
with a populated clump of RGB stars at V $\sim$ 13.0$-$13.5, and (B-V) $\sim$ 1.5.
The TO is located at V $\sim$ 14.0, (B-V) $\sim$ 0.5 mag.
Adopting the reddening derived from Fig.~8, we found that an 500 Myr isochrone
for solar metallicity nicely follows the star distribution in all the three color combination
CMDs. We estimate an apparent distance modulus
(m-M)$_V$=12.75$\pm$0.20, which translates into  a distance of 1.6$^{+0.8}_{-0.5}$ kpc from the Sun.
NGC~5715 is thus a Hyades-like cluster located in the inner disk,
similar to NGC~6583 and NGC~6404 (Carraro et al. 2005b), which apparently survived long
in spite of the difficult environment of these Galaxy regions.\\

\noindent
{\bf VdB-Hagen 164:}\\
As already underlined in Sect.~5, there is no evidence of a star concentration
neither in optical, nor in infrared  at the nominal position of this
object. 
The absence of a star cluster is also obvious when inspecting CMDs (See Fig~11, lower panels), 
since no clear sequences are seen in any of the diagrams.
We conclude then that VdB-Hagen 164 is not a star cluster, but a random
concentration of a few bright stars. 
As in the case of ESO131SC09 and NGC~5284, we, therefore,
refrain from trying any isochrone fitting.\\

\noindent
{\bf NGC 6268:}\\
This looks like  a small, compact, and young cluster. The two color diagram reveals stars of spectral type
as early as B2. Using the reddening derived from this diagram, we fit the star distribution
in the CMDs using a 150 Myr, solar metallicity, isochrone, using the same distance modulus as
for the empirical ZAMS. The TO is located at V $\approx$ 11.5, (B-V) = 0.3 and E(V-I) = 0.4 mag.
There is no indication of an RGB clump. The fit looks fine, and yields an apparent distance
modulus (m-M)$_V$= 11.4$\pm$0.2, which implies an helio-centric distance of 1.07$^{+0.20}_{-0.10}$ kpc.
The cluster is about 20 pc above the plane, and 7.1 kpc from the Galactic center.
Its position is compatible with it being part of the Carina-Sagittarius arm.\\

\noindent
{\bf Czernik 38:}\\
As commented before, we did not take U-band observations for this cluster, since it looked
to be heavily reddened. The CMDs are shown in the upper panel of Fig.~11.
The TO is located at V = 16.5, (B-V)=1.4, and (V-I) $\sim$ 1.7. 
A group of red stars at V $\sim$ 17.0 seemingly indicates the presence of a RGB clump.
We derived distance, age and reddening simultaneously by fitting a 600 Myr, solar metallicity,
isochrone, which nicely follows the star distribution in the CMD. 
This fit returns a reddening E(B-V)=1.35$\pm$0.15, and an apparent distance modulus (m-M)$_V$=15.30$\pm$0.20.
Therefore, we position Czernik~38 at 1.9$^{+0.55}_{-0.40}$ 
kpc from the Sun and at 7.2 kpc from the Galactic center,
beyond the Carina-Sagittarius arm. This location explains its significant reddening.
Like NGC~5715, Czernik~38 is thus a Hyades-like cluster located in the inner disk,
which managed to survive longer than the mean Galactic cluster lifetime.

\begin{table*}
\caption{Estimates and associated uncertainties of the fundamental parameters
of the clusters under study}
\begin{tabular}{cccccccc}
\hline
\hline
\multicolumn{1}{c}{Number} &
\multicolumn{1}{c}{Name} &
\multicolumn{1}{c}{E(B-V)}  &
\multicolumn{1}{c}{$(m-M)_V$}  &
\multicolumn{1}{c}{Age} &
\multicolumn{1}{c}{$d_{\odot}$} &
\multicolumn{1}{c}{$d_{GC}$} &
\multicolumn{1}{c}{$z$} \\
\hline
 &  & mag &  mag &  Myr & kpc & kpc & pc\\
\hline
1  & IC 2714       & 0.33$\pm$0.05 & 11.60$\pm$0.10 & 300$\pm$20  & 1.30$^{+0.10}_{-0.15}$ & 8.1 & $\sim$ 40\\
2  & NGC 4052      & 0.30$\pm$0.05 & 12.70$\pm$0.20 & 400$\pm$40  & 2.20$^{+0.50}_{-0.30}$ & 7.8 & $\sim$ 30\\
3  & ESO131SC09    &               &                &             &             &     &         \\
4  & NGC 5284      &               &                &             &             &     &          \\
5  & NGC 5316      & 0.25$\pm$0.05 & 11.50$\pm$0.20 & 100$\pm$10  & 1.40$^{+0.15}_{-0.20}$ & 7.6 & $\sim$  0\\
6  & NGC 5715      & 0.55$\pm$0.10 & 12.75$\pm$0.20 & 500$\pm$100 & 1.60$^{+0.80}_{-0.50}$ & 7.4 & $\sim$ 60\\
7  & VdB-Hagen 164 &               &                &             &             &     &          \\
8  & NGC 6268      & 0.40$\pm$0.03 & 11.40$\pm$0.20 & 150$\pm$10  & 1.07$^{+0.20}_{-0.10}$ & 7.1 & $\sim$ 20\\
9  & Czernik 38    & 1.25$\pm$0.10 & 15.30$\pm$0.20 & 600$\pm$100 & 1.90$^{+0.55}_{-0.40}$ & 7.2 & $\sim$ 80\\
\hline\hline
\end{tabular}
\end{table*}

\section{Conclusions}
We have presented and discussed UBVI photometry data for nine objects catalogued
as Galactic clusters, for which only limited, if any, data was available.
Structural parameters and fundamental properties have been derived, and are summarized in
Tables~5 and 8, respectively.\\

\noindent
The basic goal of this work was first to assess the reality of these objects using
star counts. Second, when an overdensity was detected, we looked at photometric diagrams
to probe whether  stars producing the overdensity would also exhibit
distinctive features in the TCD and CMDs.\\

\noindent
This allowed us to propose that VdB-Hagen~164 is most probably not a star cluster, since
its stars do not define either  a spatial concentration or show clear sequences in the
photometric diagrams.

We arrived at quite the same conclusion for ESO131SC09 and NGC~5284 whose stars, in spite
of showing some spatial concentration, do not produce any distinctive feature in photometric
diagrams. The appearence of a star cluster is generated by a group of bright stars, probably
physically un-correlated. 
They closely resemble the so-called open cluster remnants (Loden 1973; Pavani \& Bica 2007)
Carraro 2006), which,
in the majority of cases, are found to be random accumulations of
stars along the line of sight.\\

\noindent
The remaining targets are found to be genuine, physical, star clusters.
Their ages range from 100 to 600 million years. However, 
only two of them (NGC~5715 and NGC~6268) are very young,
and their distances are compatible with the location of the Carina-Sagittarius arm (Russeil 2003).
We can conclude they formed inside the arm and are {\it bona fide}
tracers of the arm, since with such ages they could not travel much
away from their birth-place.\\
Unfortunately, we fail to find any young cluster located beyond the Carina-Sagittarius arm in
the present sample. This clearly reflects the difficulty to penetrate the arm and to see further away
because of the high density of dust and gas, unless absorption holes allow to
detect more distant clusters (see {\it e.g.} V\'azquez et al. 1995; Baume et al. 2009).

\noindent
The oldest clusters (Czernik~38 and NGC~5715) are particularly interesting in the context
of cluster dynamical evolution and dissolution models (Lamers et al 2005), since they could survive longer
than the the typical open cluster life-time in a dense and hostile environment like
the inner disk, where tidal forces and close encounters do not
permit star clusters to survive typically more than 100-200 Myrs (Wielen 1971).\\

\noindent
Not many clusters of this age or older are known to be located at these Galacto-centric distances
(see {\it e.g.} http://www.univie.ac.at/webda/navigation.html).
This combination of age and distance is extremely useful to investigate the Galactic disk radial
abundance gradient in the inner disk (Magrini et al. 2010) and its
evolution through time.\\
Therefore, these two clusters are ideal targets for future spectroscopic follow-up to determine
their metal abundances.

\section*{Acknowledgments}
We acknowledge the staff of CTIO and LCO, in particular Edgardo Cosgrove and
Patricio Pinto, for their valuable support during the runs. The work of
A.F. Seleznev has been partly supported by the ESO Visiting Scientist Program.
We are very grateful to Sandy Strunk, who carefully revised the paper
and helped us to improve the language.
This study made use of the SIMBAD and WEBDA databases.

\end{document}